\documentclass[twocolumn,preprintnumbers,amsmath,amssymb]{revtex4}

\usepackage{graphicx}
\usepackage{bm}
\newcommand{\puw}{P_{int}}
\newcommand{\real}{{\mathrm{Re}}}
\newcommand{\etal}{\textit{et al.}}

\begin{document}

\preprint{APS/phase noise}
\title{Noise Properties of Superconducting Coplanar Waveguide Microwave Resonators }

\author{Jiansong Gao}
\affiliation{%
Physics Department, California Institute of Technology, Pasadena, CA
91125
}%
\author{Benjamin A. Mazin}%
\author{Henry G. LeDuc} %
\affiliation{ Jet Propulsion Laboratory, California Institute of
Technology, Pasadena, CA 91109
}%
\author{Jonas Zmuidzinas} %
\affiliation{%
Physics Department, California Institute of Technology, Pasadena, CA
91125
}%
\author{Peter K. Day}
\affiliation{ Jet Propulsion Laboratory, California Institute of
Technology, Pasadena, CA 91109
}%

\date{\today}

\begin{abstract}
We have measured noise in thin-film superconducting coplanar
waveguide resonators. This noise appears entirely as phase noise,
equivalent to a jitter of the resonance frequency. In contrast,
amplitude fluctuations are not observed at the sensitivity of our
measurement. The ratio between the noise power in the phase and
amplitude directions is large, in excess of 30~dB. These results
have important implications for resonant readouts of various devices
such as detectors, amplifiers, and qubits. We suggest that the phase
noise is due to two--level systems in dielectric materials.\
\end{abstract}


\maketitle

Thin-film superconducting microwave resonators are of
interest for a number of applications, including the multiplexed
readout of single electron transistors (SET)\cite{Stevenson02},
microwave kinetic inductance detectors (MKID)\cite{Day03,Mazinthesis},
normal metal-insulator-superconductor (NIS) tunnel junction detectors\cite{Schmidt03},
superconducting quantum interference devices (SQUID)\cite{Irwin04, Day06},
and qubits\cite{Wallraff04,Martinis05}.
The device to be measured
presents a variable dissipative or reactive load to the resonator,
influencing the resonator quality factor $Q_r$ or frequency $f_r$,
respectively. Changes to both $Q_r$ and $f_r$
may be determined simultaneously by sensing
the amplitude and phase of a microwave probe signal\cite{Day03}.
While several early demonstrations used hand-assembled
lumped-element circuits\cite{Stevenson02,Schmidt03,Irwin04}, frequency-domain
multiplexing of large arrays generally will require
compact microlithographed high-$Q_r$ resonators\cite{Stevenson02}.
Such resonators are also needed for strong coupling to charge
qubits\cite{Wallraff04}. Noise in microlithographed resonators has
been observed\cite{Day03,Mazinthesis} and can be a limiting factor
for device performance, but is not well understood. In this letter,
we report measurements of resonator noise, show how the noise
spectra separate into amplitude and phase components, and discuss
the physical origin of the noise.

We studied quarter-wavelength coplanar waveguide (CPW)
resonators\cite{Day03} (Fig.~1a) with center strip widths $w$ of 0.6
to 6~$\mu$m and gaps $g$ between the center strip and ground planes
of 0.4 to 4~$\mu$m, and with impedances $Z_0 \approx 50\,\Omega$.
Resonator lengths of 3 to 7~mm produce resonance frequencies $f_r$
between 4 and 10~GHz. Frequency multiplexed arrays of up to 100
resonators are coupled to a single CPW feedline. The CPW circuits
are patterned from a film of either Al ($T_c$ = 1.2~K) or Nb ($T_c$
= 9.2~K) on a crystalline substrate, either sapphire, Si or Ge. The
surfaces of the semiconductor substrates are not intentionally
oxidized, although a native oxide due to air exposure is expected to
be present.

\begin{figure}[]
\includegraphics{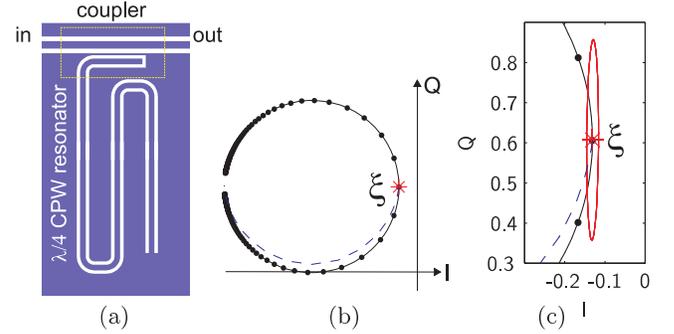}
\label{fig:IQ} \caption{(Color online) (a) Schematic illustration
(not to scale) of the resonator and feedline geometry; see Day
\etal\cite{Day03} for the equivalent circuit. Black represents the
superconducting film; white represents bare substrate. The coupler
relies on the mutual capacitance between the center strips of the
feedline and resonator CPW lines and may be considered to be a
lumped element $C = 1/ \left(f_r Z_0 \sqrt{8 \pi Q_c } \right)$
where $Q_c$ is the coupling--limited quality factor. (b) Resonance
circle of a 200~nm Nb on Si resonator at 120~mK (solid line),
quasiparticle trajectory calculated from Mattis-Bardeen theory
\cite{Mattis58}(dashed line). For this figure, the readout point
$\xi=[I, Q]$ is located at the resonance frequency $f_r$. (c) Noise
ellipse (magnified by a factor of 30). Other parameters are
$f_r$=4.35~GHz, $Q_r=3.5\times 10^5$ (coupling limited),
$w$=5~$\mu$m, $g$=1~$\mu$m, readout power $P_r \approx$ -84~dBm and
internal power $\puw \approx$-30~dBm.}
\end{figure}
A microwave synthesizer at frequency $f$ is used to excite a
resonator. The transmitted signal is amplified with a cryogenic high
electron mobility transistor (HEMT) amplifier and is compared to the
original signal using an $IQ$ mixer, whose output voltages $I$ and
$Q$ are proportional to the in-phase and quadrature amplitudes of
the transmitted signal\cite{Day03,Mazinthesis} (see Fig. 2 inset).
As $f$ is varied, the output
$\xi = [I, Q]^T$ (the superscript $T$ represents the transpose) traces out
a resonance circle (Fig.~1(b)).
With $f$ fixed, $\xi$ is seen to fluctuate about its
mean, and the fluctuations
$\delta\xi(t)=[\delta I(t), \delta Q(t)]^T$
are digitized for noise analysis, typically over a 10~s interval
using a sample rate of 250~kHz.

The fluctuations $\delta \xi(t)$ are observed to be primarily in the
direction tangent to the resonance circle, while the fluctuations in
the orthogonal direction are small. These two directions correspond
to fluctuations in the phase and amplitude of the resonator's
electric field $\vec E$, respectively. This observation can be
quantified by studying the spectral--domain noise covariance matrix
$S(\nu)$, defined by
\begin{equation}
\langle \delta\xi(\nu)\delta\xi^\dagger(\nu') \rangle =
S(\nu)\delta(\nu-\nu'),~~S(\nu) = \left(
\begin{array}{cc}
S_{II}(\nu) & S_{IQ}(\nu)\\
S^*_{IQ}(\nu) & S_{QQ}(\nu)\\
\end{array}
\right)\label{eqn:spectocov},
\end{equation}
where $\delta\xi(\nu)$ is the  Fourier transform of
the time--domain data, the dagger represents the Hermitian conjugate,
$S_{II}(\nu)$ and $S_{QQ}(\nu)$ are the auto-power spectra, and
$S_{IQ}(\nu)$ is the cross-power spectrum.
The matrix $S(\nu)$ is Hermitian and may be diagonalized
using a unitary transformation; however, we find that
the imaginary part of $S_{IQ}$ is negligible and that
an ordinary rotation applied to the real part $\real\, S(\nu)$
gives almost identical results.
The eigenvectors and  eigenvalues are calculated at every frequency,
$\nu$:
\begin{equation}
O^T(\nu)\, \real\, S(\nu)\, O(\nu) = \left(
\begin{array}{cc}\
S_{aa}(\nu) & 0 \\
0  & S_{bb}(\nu)\\
\end{array}
\right),\label{eqn:eig}
\end{equation}
where $O(\nu)=[v_a(\nu),v_b(\nu)]$ is an orthogonal rotation matrix.
We use $S_{aa}(\nu)$ and
$v_a(\nu)$ to denote the larger eigenvalue and its eigenvector.

\begin{figure}[]
\includegraphics{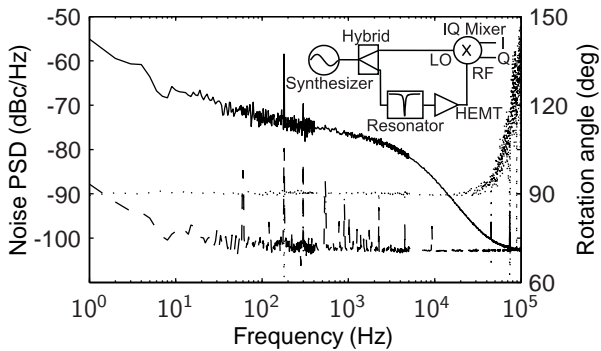}
\label{fig:specorth} \caption{Noise spectra in the phase
($S_{aa}(\nu)$, solid line) and amplitude ($S_{bb}(\nu)$, dashed
line) directions, and the rotation angle ($\phi(\nu)$, dotted line).
The noise data are from the same Nb/Si resonator under the same
condition as in Fig.~1. The inset shows the diagram of the homodyne
readout system.}
\end{figure}
A typical pair of spectra $S_{aa}(\nu)$ and $S_{bb}(\nu)$ are shown
in Fig.~2, along with the rotation angle $\phi(\nu)$, defined as the
angle between $v_a(\nu)$ and the $I$ axis. Three remarkable features
are found for all noise data. First, $\phi(\nu)$ is independent of
$\nu$ within the resonator bandwidth (the r.m.s. scatter is
$\sigma_\phi \le 0.4 ^\circ$ per 10~Hz frequency bin from 1~Hz to
5~kHz), which means that only two special directions, $v_a$ and
$v_b$, diagonalize $S(\nu)$. Eqn.~(\ref{eqn:eig}) shows that
$S_{aa}(\nu)$ and $S_{bb}(\nu)$ are the noise spectra projected into
these two constant directions. Second, $v_a$ is always
tangent to the $IQ$ resonance circle while $v_b$ is always normal to
the circle, even when $f$ is detuned from $f_r$.
Third,
$S_{aa}(\nu)$ is well above $S_{bb}(\nu)$ (see Fig.~2).
The character of the noise can be clearly visualized by plotting a
noise ellipse, defined by $ \delta \xi^T C^{-1} \delta \xi = 1$,
where $C = \int_{\nu_1}^{\nu_2}{\real\, S(\nu)d\nu}$ is the
covariance matrix for $\delta I$ and $\delta Q$ filtered for the
corresponding bandpass (we use $\nu_1 = 1$~Hz and $\nu_2 = 1$~kHz).
The major axis of the noise ellipse is always in the phase direction,
and the ratio of the two axes is always very large (Fig.~1(c)).

Fig.~2 also shows that the amplitude noise spectrum is flat except
for a $1/\nu$ knee at low frequency contributed by the electronics.
The amplitude noise is independent of whether
the synthesizer is tuned on or off the resonance, and is consistent
with the noise temperature of the HEMT amplifier.
The phase noise spectrum\cite{phasenoise} has a $1/\nu$ slope below 10~Hz,
typically a $\nu^{-1/2}$ slope above 10~Hz, and a roll-off at the
resonator bandwidth $f_r$/$2Q_r$. The phase noise is well above the
HEMT noise, usually by two or three orders of magnitude (in
rad$^2$/Hz) at low frequencies. It is well in excess of the
synthesizer phase noise contribution or the readout system
noise\cite{Day03,Mazinthesis}.

Quasiparticle fluctuations in the superconductor
can be securely ruled out as the source of the measured
noise by considering the direction in the $IQ$ plane that would
correspond to a change in quasiparticle density $\delta n_{qp}$.
Both the real and inductive parts of the complex conductivity
$\sigma$ respond linearly to $\delta n_{qp}$, $\delta \sigma =
\delta \sigma_1 - i\delta\sigma_2$, resulting in a trajectory that
is always at a non-zero angle $\psi = \tan^{-1}(\delta
\sigma_1/\delta \sigma_2)$ to the resonance circle, as indicated by
the dashed lines in Fig.~1(b) and (c). Mattis-Bardeen\cite{Mattis58}
calculations yield $\psi > 7^\circ$ for Nb below 1~K, so
quasiparticle fluctuations are strongly excluded since $\psi >>
\sigma_\phi$. Furthermore, $\psi$ is measured experimentally by
examining the response to X-ray, optical/UV, or submillimeter
photons, and is typically $\psi \approx 15^\circ$\cite{Ben06}.

The phase noise depends on the microwave power inside the resonator
($\puw$), the materials used for the resonator, and the operating
temperature. The power dependence for various material combinations
is shown in Fig.~3; all follow the scaling $S_{aa}(\nu) \propto
\puw^{-1/2}$. For comparison, amplifier phase noise is a
multiplicative effect that would give a constant noise level
independent of $\puw$, while the amplifier noise temperature is an
additive effect that would produce a $1/\puw$ dependence. Sapphire
substrates generally give lower phase noise than Si or Ge. However
the Nb/Si device showed low noise comparable with Al/sapphire,
suggesting that the etching or interface chemistry, which is
different for Nb and Al, may play a role. Two Al/Si resonators with
very different Al thicknesses and kinetic inductance
fractions\cite{Gao06b} fall onto the dashed equal-noise scaling
line, strongly suggesting that the superconductor is not responsible
for the phase noise\cite{Mazinthesis}. Furthermore, the noise of a
Nb/Si resonator decreased by a factor of 10 when warmed from 0.2~K
to 1~K, more strong evidence against superconductor noise since Nb
has $T_c$ = 9.2~K and its properties change very little for $T <<
T_c$. More detail on the temperature dependence will be
published separately.

\begin{figure}[t]
\includegraphics{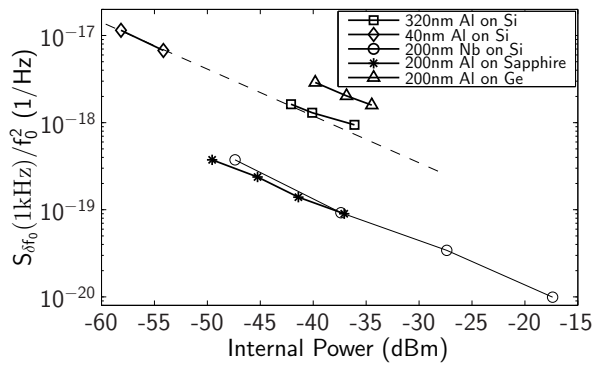}
\label{fig:noisecompare} \caption{Power and material dependence of
the phase noise at $\nu =1$~kHz. To compare resonators with
different $f_r$ and $Q_r$, phase noise is converted to fractional
frequency noise, calculated by $S_{\delta
f_r}(\nu)/f_r^2=S_{aa}(\nu)/4Q_r^2$. All the resonators have
$w$=3~$\mu$m, $g$=2~$\mu$m and are measured around 120~mK.}
\end{figure}
The evidence leads us to suggest that the noise is
caused by fluctuating two-level systems (TLS) in the dielectric
materials --- either the bulk substrate or its exposed surface, the
interface layers between the metal films and the substrate, or any
oxide layers on the metal surfaces. Models assuming a collection of
TLS with a wide range of excitation energies $E$ and relaxation
rates have long been used to explain the low temperature physical
properties of non-crystalline
solids\cite{Anderson72,Phillips72,Phillips87}. TLS are also found in
crystalline materials\cite{Kleiman87,Phillips88} but at lower
densities. Fluctuations due to TLS are of interest theoretically\cite{Yu04,Shnirman05}
and have been observed in dielectrics, either as telegraph or $1/\nu$
noise, using tunnel junctions\cite{wakai87}, single electron
transistors\cite{Zorin96}, and atomic force
microscopes\cite{Walther98}.
Other examples include telegraph noise in the resistance of
metallic nanoconstrictions\cite{Ralls88}
and qubit dephasing effects\cite{Martinis05}.

Recent experiments\cite{Wallraff04,schuster05} have illustrated
the reactive loading effect that occurs when a single TLS (a qubit) is
coupled to a microwave resonator. For weak coupling $g$ or large
detuning $\Delta f=|E/h-f_r|$, $g << \Delta f$, this reactive loading
causes the  resonator frequency to shift by
$\pm g^2/\Delta f$ depending on the quantum state of the TLS.
Thermal fluctuations -- absorption or emission of thermal phonons
by a collection of TLS -- could therefore cause phase noise,
equivalent to a fluctuating dielectric constant
$\delta \epsilon(\vec r, t)$.
For this model one expects the noise to vanish exponentially for
$T << h f_r / 2 k$ as the TLS settle into their ground states,
while the increase in TLS--phonon transition rates
might explain the observed noise decrease at high temperatures.
However, at present we do not have a model which
quantitatively explains the data.
Alternatively, TLS frequency fluctuations
$\delta E(t)/h$ produced by TLS-TLS
interactions\cite{Black77,Phillips87} and observed in single
molecule optical fluorescence experiments\cite{Ambrose91,Boiron99}
should give phase noise with a power--law rather than exponential
decrease at low temperatures.

The $\puw^{-1/2}$ noise scaling is indicative of TLS saturation;
otherwise, the phase noise would be independent of $\puw$ as
expected for dielectric constant fluctuations.
Similarly, power-independent thermal fluctuations of the TLS
dielectric polarization $\delta \vec P$ \cite{Shnirman05} are ruled
out since this would give additive noise with equal amplitude
and phase components and scaling as $1/\puw$.
TLS saturation effects are well known\cite{Phillips87,Martinis05};
indeed, $f_r$ and $Q_r$ show anomalous temperature
and power dependence\cite{Rami06} for $T<<T_c$.
Saturation of a single TLS depends on its frequency
detuning $\Delta f$, its position $\vec r$ in the spatially
varying resonator field $\vec E(\vec r)$, and the orientation
and strength of its  dipole moment $\vec d$.
The growth of the saturation zone in this parameter space
with increasing $\puw$ may explain the observed $\puw^{-1/2}$
noise scaling.

The microwave fields in our resonators are
sufficiently strong to cause TLS saturation.
In the Bloch equation
framework, saturation for zero detuning occurs when
$\Omega^2 T_1 T_2 >1$, where $\Omega = \vec d \cdot \vec E / \hbar$
is the Rabi frequency and $T_1^{-1}$, $T_2^{-1}$ are the usual
energy relaxation and dephasing rates\cite{Wilhelm06}. The
distribution of $T_1^{-1}$ is extremely broad\cite{Phillips87},
extending above and below the range of noise frequencies we observe,
because $T_1^{-1} \propto \Delta_0^2 / E^2$ is controlled by the
tunnel coupling $\Delta_0$ and is therefore exponentially sensitive
to the potential barrier\cite{Phillips87,Wilhelm06}. For silica,
$T_1^{-1} \approx 1~\mathrm{MHz}\, (E/4\, \mathrm{GHz})^3\
(\Delta_0/E)^2 \coth(E/2kT)$. The dipole moment $\vec d$ is
proportional to $\Delta_0/E$\cite{Phillips87}, so $\Omega^2 T_1$ is
independent of $\Delta_0$. Meanwhile, $T_2^{-1}$ scales as
$(1-\Delta_0^2/E^2)$\cite{Phillips87,Wilhelm06} and has a relatively
narrow distribution; for silica, $T_2^{-1} \approx 10
\,$MHz$\,(T/200\, \mathrm{mK})^{1.5}$\cite{putikka87}. Saturation is
therefore mostly independent of $\Delta_0$, and occurs when $\vec E
\cdot \vec d / |\vec d|$ exceeds a critical value $E_c(f_r, T)$.
Indeed, the shape of the observed
phase noise spectrum does not change significantly with $\puw$ until
the superconductors become nonlinear\cite{Mazinthesis}.
For silica,
$E_c(4~\mathrm{GHz}, 200~\mathrm{mK}) \approx 31\, \mathrm{V\, m^{-1}}$
and corresponds to $\puw \approx -75$~dBm which is significantly below
the power levels we use (see Fig.~3).
We also find
$E_c(7.2~\mathrm{GHz}, 25~\mathrm{mK}) \approx 10\, \mathrm{V\, m^{-1}}$,
which is appropriate for the experimental conditions of
Martinis \etal\cite{Martinis05}
and is similar to their observed onset of TLS saturation.

In summary, we find that lithographed resonators display
$\nu^{-1/2}$ noise directed purely in the phase direction
which varies with readout power, temperature, and
the substrate material.
These results are important for the device optimization --- devices
relying on resistive loading should be able to avoid this noise
source.
For MKID detectors, the $\sin \psi$ amplitude component of the
signal is already available when using an $IQ$ readout, and an
optimally weighted phase/amplitude measurement can substantially
improve the sensitivity at low frequencies where the phase noise is
larger.

We thank Sunil Golwala, Kent Irwin, Andrew Lange, Konrad Lehnert,
John Martinis and Harvey Moseley for useful discussions. This work
was supported in part by the NASA Science Mission Directorate, JPL,
Gordon and Betty Moore Foundation, and Alex Lidow, a Caltech
Trustee.


\begin{thebibliography}{30}
\expandafter\ifx\csname
natexlab\endcsname\relax\def\natexlab#1{#1}\fi
\expandafter\ifx\csname bibnamefont\endcsname\relax
  \def\bibnamefont#1{#1}\fi
\expandafter\ifx\csname bibfnamefont\endcsname\relax
  \def\bibfnamefont#1{#1}\fi
\expandafter\ifx\csname citenamefont\endcsname\relax
  \def\citenamefont#1{#1}\fi
\expandafter\ifx\csname url\endcsname\relax
  \def\url#1{\texttt{#1}}\fi
\expandafter\ifx\csname urlprefix\endcsname\relax\def\urlprefix{URL
}\fi \providecommand{\bibinfo}[2]{#2}
\providecommand{\eprint}[2][]{\url{#2}}

\bibitem[{\citenamefont{{Stevenson} et~al.}(2002)\citenamefont{{Stevenson},
  {Pellerano}, {Stahle}, {Aidala}, and {Schoelkopf}}}]{Stevenson02}
\bibinfo{author}{\bibfnamefont{T.~R.} \bibnamefont{{Stevenson}}},
  \bibinfo{author}{\bibfnamefont{F.~A.} \bibnamefont{{Pellerano}}},
  \bibinfo{author}{\bibfnamefont{C.~M.} \bibnamefont{{Stahle}}},
  \bibinfo{author}{\bibfnamefont{K.}~\bibnamefont{{Aidala}}}, \bibnamefont{and}
  \bibinfo{author}{\bibfnamefont{R.~J.} \bibnamefont{{Schoelkopf}}},
  \bibinfo{journal}{Applied Physics Letters} \textbf{\bibinfo{volume}{80}},
  \bibinfo{pages}{3012} (\bibinfo{year}{2002}).

\bibitem[{\citenamefont{Day et~al.}(2003)\citenamefont{Day, LeDuc, Mazin,
  Vayonakis, and Zmuidzinas}}]{Day03}
\bibinfo{author}{\bibfnamefont{P.~K.} \bibnamefont{Day}},
  \bibinfo{author}{\bibfnamefont{H.~G.} \bibnamefont{LeDuc}},
  \bibinfo{author}{\bibfnamefont{B.~A.} \bibnamefont{Mazin}},
  \bibinfo{author}{\bibfnamefont{A.}~\bibnamefont{Vayonakis}},
  \bibnamefont{and}
  \bibinfo{author}{\bibfnamefont{J.}~\bibnamefont{Zmuidzinas}},
  \bibinfo{journal}{Nature} \textbf{\bibinfo{volume}{425}},
  \bibinfo{pages}{817} (\bibinfo{year}{2003}).

\bibitem[{\citenamefont{Mazin}(2004)}]{Mazinthesis}
\bibinfo{author}{\bibfnamefont{B.~A.} \bibnamefont{Mazin}}, Ph.D. thesis,
  \bibinfo{school}{Caltech} (\bibinfo{year}{2004}).

\bibitem[{\citenamefont{Schmidt et~al.}(2003)\citenamefont{Schmidt, Yung, and
  Cleland}}]{Schmidt03}
\bibinfo{author}{\bibfnamefont{D.~R.} \bibnamefont{Schmidt}},
  \bibinfo{author}{\bibfnamefont{C.~S.} \bibnamefont{Yung}}, \bibnamefont{and}
  \bibinfo{author}{\bibfnamefont{A.~N.} \bibnamefont{Cleland}},
  \bibinfo{journal}{Appl. Phys. Lett.} \textbf{\bibinfo{volume}{83}},
  \bibinfo{pages}{1002} (\bibinfo{year}{2003}).

\bibitem[{\citenamefont{Irwin and Lehnert}(2004)}]{Irwin04}
\bibinfo{author}{\bibfnamefont{K.~D.} \bibnamefont{Irwin}} \bibnamefont{and}
  \bibinfo{author}{\bibfnamefont{K.~W.} \bibnamefont{Lehnert}},
  \bibinfo{journal}{Appl. Phys. Lett.} \textbf{\bibinfo{volume}{85}},
  \bibinfo{pages}{2107} (\bibinfo{year}{2004}).

\bibitem[{\citenamefont{Hahn et~al.}(2006)\citenamefont{Hahn, Bumble, Leduc,
  Weilert, and Day}}]{Day06}
\bibinfo{author}{\bibfnamefont{I.}~\bibnamefont{Hahn}},
  \bibinfo{author}{\bibfnamefont{B.}~\bibnamefont{Bumble}},
  \bibinfo{author}{\bibfnamefont{H.}~\bibnamefont{Leduc}},
  \bibinfo{author}{\bibfnamefont{M.}~\bibnamefont{Weilert}}, \bibnamefont{and}
  \bibinfo{author}{\bibfnamefont{P.}~\bibnamefont{Day}}, in
  \emph{\bibinfo{booktitle}{AIP Conference proceedings:24th International
  Conference on Low Temperature Physics}} (\bibinfo{year}{2006}), vol.
  \bibinfo{volume}{850}, p. \bibinfo{pages}{1613}.

\bibitem[{\citenamefont{Wallraff et~al.}(2004)\citenamefont{Wallraff, Schuster,
  Blais, Frunzio, Huang, Majer, Kumar1, Girvin, and Schoelkopf}}]{Wallraff04}
\bibinfo{author}{\bibfnamefont{A.}~\bibnamefont{Wallraff}},
  \bibinfo{author}{\bibfnamefont{D.~I.} \bibnamefont{Schuster}},
  \bibinfo{author}{\bibfnamefont{A.}~\bibnamefont{Blais}},
  \bibinfo{author}{\bibfnamefont{L.}~\bibnamefont{Frunzio}},
  \bibinfo{author}{\bibfnamefont{R.-S.} \bibnamefont{Huang}},
  \bibinfo{author}{\bibfnamefont{J.}~\bibnamefont{Majer}},
  \bibinfo{author}{\bibfnamefont{S.}~\bibnamefont{Kumar1}},
  \bibinfo{author}{\bibfnamefont{S.~M.} \bibnamefont{Girvin}},
  \bibnamefont{and} \bibinfo{author}{\bibfnamefont{R.~J.}
  \bibnamefont{Schoelkopf}}, \bibinfo{journal}{Nature}
  \textbf{\bibinfo{volume}{431}}, \bibinfo{pages}{162} (\bibinfo{year}{2004}).

\bibitem[{\citenamefont{Martinis et~al.}(2005)\citenamefont{Martinis, Cooper,
  McDermott, Steffen, Ansmann, Osborn, Cicak, Oh, Pappas, Simmonds, and Yu}}]{Martinis05}
\bibinfo{author}{\bibfnamefont{J.~M.} \bibnamefont{Martinis}},
  \bibinfo{author}{\bibfnamefont{K.~B.} \bibnamefont{Cooper}},
  \bibinfo{author}{\bibfnamefont{R.}~\bibnamefont{McDermott}},
  \bibinfo{author}{\bibfnamefont{M.}~\bibnamefont{Steffen}},
  \bibinfo{author}{\bibfnamefont{M.}~\bibnamefont{Ansmann}},
  \bibinfo{author}{\bibfnamefont{K.~D.} \bibnamefont{Osborn}},
  \bibinfo{author}{\bibfnamefont{K.}~\bibnamefont{Cicak}},
  \bibinfo{author}{\bibfnamefont{S.}~\bibnamefont{Oh}},
  \bibinfo{author}{\bibfnamefont{D.~P.} \bibnamefont{Pappas}},
  \bibinfo{author}{\bibfnamefont{R.~W.} \bibnamefont{Simmonds}},
  \bibnamefont{and} \bibinfo{author}{\bibfnamefont{C.~C.} \bibnamefont{Yu}},
  \bibinfo{journal}{Physical Review Letters}
  \textbf{\bibinfo{volume}{95}}, \bibinfo{eid}{210503} (\bibinfo{year}{2005}).

\bibitem[{pha()}]{phasenoise}
\bibinfo{note}{The phase noise spectra we present are conceptually different
  from those commonly used to represent the phase noise of active oscillators
  or frequency sources. Our noise spectra refer to a passive resonator driven
  at fixed frequency; in this case, a resonance frequency shift $\delta f_r$
  causes a proportional phase shift $\delta \theta = 2 Q_r \delta f_r/f_r$ in
  the resonator field. In contrast, if we locked an active oscillator to our
  passive resonator, the oscillator's output phase shift relative to a perfect
  clock would now be the time integral of $2 \pi\, \delta f_r(t)$, and we would
  therefore observe an extra $1/\nu^2$ factor in the oscillator's phase noise
  spectrum at low frequencies.}

\bibitem[{\citenamefont{Mattis and Bardeen}(1958)}]{Mattis58}
\bibinfo{author}{\bibfnamefont{D.~C.} \bibnamefont{Mattis}} \bibnamefont{and}
  \bibinfo{author}{\bibfnamefont{J.}~\bibnamefont{Bardeen}},
  \bibinfo{journal}{Phys. Rev.} \textbf{\bibinfo{volume}{111}},
  \bibinfo{pages}{412} (\bibinfo{year}{1958}).

\bibitem[{\citenamefont{Mazin et~al.}(2006)\citenamefont{Mazin, Bumble, Day,
  Eckart, Golwala, Zmuidzinas, and Harrison}}]{Ben06}
\bibinfo{author}{\bibfnamefont{B.~A.} \bibnamefont{Mazin}},
  \bibinfo{author}{\bibfnamefont{B.}~\bibnamefont{Bumble}},
  \bibinfo{author}{\bibfnamefont{P.~K.} \bibnamefont{Day}},
  \bibinfo{author}{\bibfnamefont{M.~E.} \bibnamefont{Eckart}},
  \bibinfo{author}{\bibfnamefont{S.}~\bibnamefont{Golwala}},
  \bibinfo{author}{\bibfnamefont{J.}~\bibnamefont{Zmuidzinas}},
  \bibnamefont{and} \bibinfo{author}{\bibfnamefont{F.~A.}
  \bibnamefont{Harrison}}, \bibinfo{journal}{Appl. Phys. Lett.}
  \textbf{\bibinfo{volume}{89}}, \bibinfo{eid}{222507} (\bibinfo{year}{2006}).

\bibitem[{\citenamefont{{Gao} et~al.}(2006)\citenamefont{{Gao}, {Zmuidzinas},
  {Mazin}, {Day}, and {Leduc}}}]{Gao06b}
\bibinfo{author}{\bibfnamefont{J.}~\bibnamefont{{Gao}}},
  \bibinfo{author}{\bibfnamefont{J.}~\bibnamefont{{Zmuidzinas}}},
  \bibinfo{author}{\bibfnamefont{B.~A.} \bibnamefont{{Mazin}}},
  \bibinfo{author}{\bibfnamefont{P.~K.} \bibnamefont{{Day}}}, \bibnamefont{and}
  \bibinfo{author}{\bibfnamefont{H.~G.} \bibnamefont{{Leduc}}},
  \bibinfo{journal}{Nucl. Instrum. Methods Phys. Res., Sect. A}
  \textbf{\bibinfo{volume}{559}}, \bibinfo{pages}{585} (\bibinfo{year}{2006}).

\bibitem[{\citenamefont{Anderson et~al.}(1972)\citenamefont{Anderson, Halperin,
  and Varma}}]{Anderson72}
\bibinfo{author}{\bibfnamefont{P.~W.} \bibnamefont{Anderson}},
  \bibinfo{author}{\bibfnamefont{B.~I.} \bibnamefont{Halperin}},
  \bibnamefont{and} \bibinfo{author}{\bibfnamefont{C.~M.} \bibnamefont{Varma}},
  \bibinfo{journal}{Phil. Mag.} \textbf{\bibinfo{volume}{25}},
  \bibinfo{pages}{1} (\bibinfo{year}{1972}).

\bibitem[{\citenamefont{Phillips}(1972)}]{Phillips72}
\bibinfo{author}{\bibfnamefont{W.~A.} \bibnamefont{Phillips}},
  \bibinfo{journal}{J. Low Temp. Phys.} \textbf{\bibinfo{volume}{7}},
  \bibinfo{pages}{351} (\bibinfo{year}{1972}).

\bibitem[{\citenamefont{Phillips}(1987)}]{Phillips87}
\bibinfo{author}{\bibfnamefont{W.~A.} \bibnamefont{Phillips}},
  \bibinfo{journal}{Rep. Prog. Phys.} \textbf{\bibinfo{volume}{50}},
  \bibinfo{pages}{1657} (\bibinfo{year}{1987}).

\bibitem[{\citenamefont{Kleiman et~al.}(1987)\citenamefont{Kleiman, Agnolet,
  and Bishop}}]{Kleiman87}
\bibinfo{author}{\bibfnamefont{R.~N.} \bibnamefont{Kleiman}},
  \bibinfo{author}{\bibfnamefont{G.}~\bibnamefont{Agnolet}}, \bibnamefont{and}
  \bibinfo{author}{\bibfnamefont{D.~J.} \bibnamefont{Bishop}},
  \bibinfo{journal}{Phys. Rev. Lett.} \textbf{\bibinfo{volume}{59}},
  \bibinfo{pages}{2079} (\bibinfo{year}{1987}).

\bibitem[{\citenamefont{Phillips}(1988)}]{Phillips88}
\bibinfo{author}{\bibfnamefont{W.~A.} \bibnamefont{Phillips}},
  \bibinfo{journal}{Phys. Rev. Lett.} \textbf{\bibinfo{volume}{61}},
  \bibinfo{pages}{2632} (\bibinfo{year}{1988}).

\bibitem[{\citenamefont{Yu}(2004)}]{Yu04}
\bibinfo{author}{\bibfnamefont{C.~C.} \bibnamefont{Yu}}, \bibinfo{journal}{J.
  Low Temp. Phys.} \textbf{\bibinfo{volume}{137}}, \bibinfo{pages}{251}
  (\bibinfo{year}{2004}).

\bibitem[{\citenamefont{Shnirman et~al.}(2005)\citenamefont{Shnirman, Schon,
  Martin, and Makhlin}}]{Shnirman05}
\bibinfo{author}{\bibfnamefont{A.}~\bibnamefont{Shnirman}},
  \bibinfo{author}{\bibfnamefont{G.}~\bibnamefont{Schon}},
  \bibinfo{author}{\bibfnamefont{I.}~\bibnamefont{Martin}}, \bibnamefont{and}
  \bibinfo{author}{\bibfnamefont{Y.}~\bibnamefont{Makhlin}},
  \bibinfo{journal}{Phys. Rev. Lett.} \textbf{\bibinfo{volume}{94}},
  \bibinfo{eid}{127002} (\bibinfo{year}{2005}).

\bibitem[{\citenamefont{Wakai and Van~Harlingen}(1987)}]{wakai87}
\bibinfo{author}{\bibfnamefont{R.~T.} \bibnamefont{Wakai}} \bibnamefont{and}
  \bibinfo{author}{\bibfnamefont{D.~J.} \bibnamefont{Van~Harlingen}},
  \bibinfo{journal}{Phys. Rev. Lett.} \textbf{\bibinfo{volume}{58}},
  \bibinfo{pages}{1687} (\bibinfo{year}{1987}).

\bibitem[{\citenamefont{Zorin et~al.}(1996)\citenamefont{Zorin, Ahlers,
  Niemeyer, Weimann, and Wolf}}]{Zorin96}
\bibinfo{author}{\bibfnamefont{A.~B.} \bibnamefont{Zorin}},
  \bibinfo{author}{\bibfnamefont{F.-J.} \bibnamefont{Ahlers}},
  \bibinfo{author}{\bibfnamefont{J.}~\bibnamefont{Niemeyer}},
  \bibinfo{author}{\bibfnamefont{T.}~\bibnamefont{Weimann}}, \bibnamefont{and}
  \bibinfo{author}{\bibfnamefont{H.}~\bibnamefont{Wolf}},
  \bibinfo{journal}{Phys. Rev. B} \textbf{\bibinfo{volume}{53}},
  \bibinfo{pages}{13682} (\bibinfo{year}{1996}).

\bibitem[{\citenamefont{Walther et~al.}(1998)\citenamefont{Walther,
  Vidal~Russel, Israeloff, and Alvarez~Gomariz}}]{Walther98}
\bibinfo{author}{\bibfnamefont{L.}~\bibnamefont{Walther}},
  \bibinfo{author}{\bibfnamefont{E.}~\bibnamefont{Vidal~Russel}},
  \bibinfo{author}{\bibfnamefont{N.}~\bibnamefont{Israeloff}},
  \bibnamefont{and}
  \bibinfo{author}{\bibfnamefont{H.}~\bibnamefont{Alvarez~Gomariz}},
  \bibinfo{journal}{Appl. Phys. Lett.} \textbf{\bibinfo{volume}{72}},
  \bibinfo{pages}{3223} (\bibinfo{year}{1998}).

\bibitem[{\citenamefont{Ralls and Buhrman}(1988)}]{Ralls88}
\bibinfo{author}{\bibfnamefont{K.~S.} \bibnamefont{Ralls}} \bibnamefont{and}
  \bibinfo{author}{\bibfnamefont{R.~A.} \bibnamefont{Buhrman}},
  \bibinfo{journal}{Phys. Rev. Lett.} \textbf{\bibinfo{volume}{60}},
  \bibinfo{pages}{2434} (\bibinfo{year}{1988}).

\bibitem[{\citenamefont{Schuster et~al.}(2005)\citenamefont{Schuster, Wallraff,
  Blais, Frunzio, Huang, Majer, Girvin, and Schoelkopf}}]{schuster05}
\bibinfo{author}{\bibfnamefont{D.~I.} \bibnamefont{Schuster}},
  \bibinfo{author}{\bibfnamefont{A.}~\bibnamefont{Wallraff}},
  \bibinfo{author}{\bibfnamefont{A.}~\bibnamefont{Blais}},
  \bibinfo{author}{\bibfnamefont{L.}~\bibnamefont{Frunzio}},
  \bibinfo{author}{\bibfnamefont{R.-S.} \bibnamefont{Huang}},
  \bibinfo{author}{\bibfnamefont{J.}~\bibnamefont{Majer}},
  \bibinfo{author}{\bibfnamefont{S.~M.} \bibnamefont{Girvin}},
  \bibnamefont{and} \bibinfo{author}{\bibfnamefont{R.~J.}
  \bibnamefont{Schoelkopf}}, \bibinfo{journal}{Phys. Rev. Lett.}
  \textbf{\bibinfo{volume}{94}}, \bibinfo{eid}{123602} (\bibinfo{year}{2005}).

\bibitem[{\citenamefont{Black and Halperin}(1977)}]{Black77}
\bibinfo{author}{\bibfnamefont{J.~L.} \bibnamefont{Black}} \bibnamefont{and}
  \bibinfo{author}{\bibfnamefont{B.~I.} \bibnamefont{Halperin}},
  \bibinfo{journal}{Phys. Rev. B} \textbf{\bibinfo{volume}{16}},
  \bibinfo{pages}{2879} (\bibinfo{year}{1977}).

\bibitem[{\citenamefont{Ambrose and Moerner}(1991)}]{Ambrose91}
\bibinfo{author}{\bibfnamefont{W.~P.} \bibnamefont{Ambrose}} \bibnamefont{and}
  \bibinfo{author}{\bibfnamefont{W.~E.} \bibnamefont{Moerner}},
  \bibinfo{journal}{Nature} \textbf{\bibinfo{volume}{349}},
  \bibinfo{pages}{225} (\bibinfo{year}{1991}).

\bibitem[{\citenamefont{Boiron et~al.}(1999)\citenamefont{Boiron, Tamarat,
  Lounis, Brown, and Orrit}}]{Boiron99}
\bibinfo{author}{\bibfnamefont{A.~M.} \bibnamefont{Boiron}},
  \bibinfo{author}{\bibfnamefont{P.}~\bibnamefont{Tamarat}},
  \bibinfo{author}{\bibfnamefont{B.}~\bibnamefont{Lounis}},
  \bibinfo{author}{\bibfnamefont{R.}~\bibnamefont{Brown}}, \bibnamefont{and}
  \bibinfo{author}{\bibfnamefont{M.}~\bibnamefont{Orrit}},
  \bibinfo{journal}{Chem. Phys.} \textbf{\bibinfo{volume}{247}},
  \bibinfo{pages}{119} (\bibinfo{year}{1999}).

\bibitem[{\citenamefont{Barends et~al.}(2006)\citenamefont{Barends, Baselmans,
  Hovenier, Gao, Yates, Klapwijk, and Hoevers}}]{Rami06}
\bibinfo{author}{\bibfnamefont{R.}~\bibnamefont{Barends}},
  \bibinfo{author}{\bibfnamefont{J.~J.~A.} \bibnamefont{Baselmans}},
  \bibinfo{author}{\bibfnamefont{J.~N.} \bibnamefont{Hovenier}},
  \bibinfo{author}{\bibfnamefont{J.~R.} \bibnamefont{Gao}},
  \bibinfo{author}{\bibfnamefont{S.~J.~C.} \bibnamefont{Yates}},
  \bibinfo{author}{\bibfnamefont{T.~M.} \bibnamefont{Klapwijk}},
  \bibnamefont{and} \bibinfo{author}{\bibfnamefont{H.~F.~C.}
  \bibnamefont{Hoevers}}, \bibinfo{journal}{IEEE Trans. Appl. Supercond}
  (\bibinfo{year}{2006}), \bibinfo{note}{submitted}.

\bibitem[{\citenamefont{Wilhelm et~al.}(2006)\citenamefont{Wilhelm, Storcz,
  Hartmann, and Geller}}]{Wilhelm06}
\bibinfo{author}{\bibfnamefont{F.~K.} \bibnamefont{Wilhelm}},
  \bibinfo{author}{\bibfnamefont{M.~J.} \bibnamefont{Storcz}},
  \bibinfo{author}{\bibfnamefont{U.}~\bibnamefont{Hartmann}}, \bibnamefont{and}
  \bibinfo{author}{\bibfnamefont{M.~R.} \bibnamefont{Geller}}
  (\bibinfo{year}{2006}), \bibinfo{note}{arXiv:cond--mat/0603637}.

\bibitem[{\citenamefont{Putikka and Huber}(1987)}]{putikka87}
\bibinfo{author}{\bibfnamefont{W.~O.} \bibnamefont{Putikka}} \bibnamefont{and}
  \bibinfo{author}{\bibfnamefont{D.~L.} \bibnamefont{Huber}},
  \bibinfo{journal}{Phys. Rev. B} \textbf{\bibinfo{volume}{36}},
  \bibinfo{pages}{3436} (\bibinfo{year}{1987}).

\end{thebibliography}

%
%

\end{document}